\documentclass[12pt]{article}
\begin{document}

\vskip 0.1cm

\centerline{\large \bf HIDDEN SYMMETRY AND SEPARATION OF VARIABLES}
\centerline{\large \bf IN THE PROBLEM OF TWO CENTRES}
\centerline{\large \bf WITH A CONFINEMENT-TYPE POTENTIAL}

\vskip 0.7cm

\centerline{V.Yu.Lazur$^{\ast}$, V.M.Dobosh, V.V.Rubish$^{\ddagger}$, M.D.Melika}

\vskip .2cm

\centerline{\sl Uzhgorod State University,}
\centerline{\sl Department of Theoretical Physics, Voloshin str. 32,}
\centerline{\sl 88000 Uzhgorod, Ukraine}

\vskip 0.7cm

\vskip 1.0cm

\begin{abstract}
An additional spheroidal integral of motion and a group of dynamic
symmetry in a model quantum-mechanical problem of two centres
$eZ_{1}Z_{2}\omega $ with Coulomb and oscillator interactions is
obtained, the group properties of its solutions being studied.
$P\left( 3\right) \otimes P\left(2,1\right) $, $P\left( 5,1\right)
$ and $P\left( 4,2\right) $ groups are considered as the dynamic
symmetry groups of the problem, among them $P\left( 3\right)
\otimes P\left( 2,1\right) $ group possessing the smallest number
of parameters. The obtained results may appear useful at the
calcuations of QQq-baryons and QQg-mesons energy spectra.
\end{abstract}

\vskip .3cm

\vskip 10cm

\hrule

\vskip .3cm

\noindent
\vfill $ \begin{array}{ll} ^{\ast}\mbox{{\it e-mail
address:}} &
 \mbox{lazur@univ.uzhgorod.ua}
\end{array}
$

$ \begin{array}{ll} ^{\ddagger}\mbox{{\it e-mail address:}} &
 \mbox{vrubish@univ.uzhgorod.ua}
\end{array}
$

\vfill

\baselineskip=14pt

\section{Introduction}

As a rule, when systems, possessing hidden (or higher dynamic) symmetry, are
considered, two methods are used \cite{1,2}. The first of them consists in
rewriting Schroedinger equation and putting it in the form where the
symmetry, having been hidden before, becomes explicit. The second one
implies the construction of integrals of motion which play the role of
hidden symmetry group generators.

In the proposed paper, based on the example of a physically important model
of confinement-type two-centred potential we try to emphasize the deep
relationship of the hidden symmetry to the possibility of separation of
variables in the Schroedinger equation. The awareness of such kind of
relationships in two recent decades \cite{3} has resulted in the intense
application of the method of separation of variables to the equations of
mathematical physics and led to a series of important and far from trivial
results in this field of mathematics (See, for instance, \cite{4,5}). The
method of separation of variables is much simpler compared to the two above
methods. In the framework of this method the eigenvalues of the hidden
symmetry group generators acquire the sense of separation constants, and the
eigenfunctions, common for the Hamiltonian and the generators, which
commutate with the Hamiltonian and with each other, are the solutions of the
Schroedinger equation in the corresponding coordinates.

Below, using the separation of variables, the group properties of a model
quantum problem of the motion of a light particle (a gluon) in the field of
two heavy particles (a quark-antiquark pair) are studied. Recently this
problem has become the subject of intense studies due to its relation to a
wide range of problems of hadron physics: models of baryons with two heavy
quarks (QQq barions) \cite{6} and models of heavy hybride mesons with open
flavor (QQg mesons) \cite{7}. In spite of the lack of strict theoretical
substantiation, the potential models give a satisfactory description of mass
spectra for heavy mesons and baryons (See e.g. \cite{6}-\cite{8} and
references therein), which, according to modern views, are bound states of
quarks. While modelling the interquark interaction potential, as a rule,
confinement-type potentials are used \cite{8}-\cite{9}. One of such
potentials is a so-called Cornell potential, containing a Coulomb-like term
of single-gluon exchange and a term, responsible for the string interaction,
providing the quark confinement. The confinement part of the potential is
most often modelled by a spatial spherically symmetrical oscillator
potential \cite{6,7}. Then in a non-relativistic approximation the motion of
a light quark (gluon) in the field of two heavy quarks can be described by a
stationary Schroedinger equation with a model combined potential, being the
sum of the potential of two Coulomb centres and the potential of two
harmonic oscillators:

\begin{equation}
V\left( r_{1},r_{2}\right) =-\frac{Z_{1}}{r_{1}}-\frac{Z_{2}}{r_{2}}+\omega
^{2}\left( r_{1}^{2}+r_{2}^{2}\right)  \label{f1}
\end{equation}

In this formula $r_{1}$ and $r_{2}$ are the distances from the particle to
the fixed force centres 1 and 2, $Z_{1,2}=\frac{2}{3}\alpha _{s}$, $\alpha
_{s}-$ the strong interaction constant, and the phenomenological parameter $%
\omega$ is chosen from the condition of the best agreement of the calculated
mass spectra of the quark system with the experimental data. In order to
avoid ambiguities, one should mention that in our consideration, concerning
not only the case of purely Coulomb interaction of the light particle with
each of the centres, the notion of the force centre is preserved for the $%
r_{1,2}=0$ points, where the combined potential (\ref{f1}) has the
singularities.

In the dimensionless variables the Schroedinger equation with the model
potential (\ref{f1}) is given by
\begin{equation}
\widehat{H}\Psi \equiv \left[ -\frac{1}{2}\bigtriangleup -\frac{Z_{1}}{r_{1}}%
-\frac{Z_{2}}{r_{2}}+\omega ^{2}\left( r_{1}^{2}+r_{2}^{2}\right) \right]
\Psi \left( \overrightarrow{r};R\right) =E\left( R\right) \Psi \left(
\overrightarrow{r};R\right)   \label{f2}
\end{equation}
where $r$ is the distance from the particle to the midpoint of the
intercentre distance $R$, $E\left( R\right) $ and $\Psi \left(
\overrightarrow{r};R\right) $ are the particle energy and wave function.
Hereinafter the spectral problem for the Schroedinger equation (\ref{f2})
with the combined potential (\ref{f1}) is conveniently denoted by $%
eZ_{1}Z_{2}\omega $. The sense of such notation follows from the fact that
traditional quantum-mechanical problem of two purely Coulomb centres \cite
{10} has a standard notation $eZ_{1}Z_{2}$. Note that the Schroedinger
equation for the $eZ_{1}Z_{2}$ problem can be obtained from the equation (%
\ref{f2}) by a limiting transition $\omega \rightarrow 0$.

The group properties, eigenvalue and eigenfunction spectrum for the $%
eZ_{1}Z_{2}$ problem of two Coulomb centres have been studied substantially
\cite{10}-\cite{16}. Namely, the choice of a certain non-canonical basis in
a group being a direct product of two groups of motions of three-dimensional
spaces $P\left( 3\right) \otimes P\left( 2,1\right) $, or in wider groups of
motions of six-dimensional spaces $P\left( 5,1\right) $ and $P\left(
4,2\right) $ is known to result in the necessity of the problem, equivalent
to $eZ_{1}Z_{2}$, to be solved. The consequence of these group properties of
$eZ_{1}Z_{2}$ solutions problem is a linear algebra of two-centred
integrals, obtained in \cite{16}.

Here we show that for our case of $eZ_{1}Z_{2}\omega $ problem a generally
similar situation takes place. This problem can also be considered as a
problem of theory of representations of certain non-compact groups, where
the function being a product of a quasiradial and a quasiangular two-centred
functions by $\exp \left( im\alpha +i\widetilde{m}\beta \right) $, comprises
the basis of a degenerate non-canonical representation of the group being a
direct product of two three-dimentional space motion groups $P\left(
3\right) \otimes P\left( 2,1\right) $, or wider six-dimensional space morion
groups $P\left( 5,1\right) $, $P\left( 4,2\right) $ etc.

The term ''non-canonical representation'' is used in the present paper, like
in \cite{13}-\cite{15}, for less studied representations where not all the
operators of the complete set of the observed ones are the invariants of the
subgroups of the considered group.

\section{Spheroidal integral of motion in the problem}

The variables in Eq. (\ref{f2}) can be separated by introducing an elongated
spheroidal (elliptical) coordinate system $\left\{ \xi \eta \alpha \right\} $
with the origin in the midpoint of $R$ segment and foci in its endpoints
\cite{10}:
\begin{equation}
\begin{array}{cc}
\xi =\left( r_{1}+r_{2}\right) /R, & 1\leq \xi <\infty , \\
\eta =\left( r_{1}-r_{2}\right) /R, & -1\leq \eta \leq 1, \\
\alpha =\arctan \left( \frac{x_{2}}{x_{1}}\right) , & 0\leq \alpha <2\pi .
\end{array}
\label{f3}
\end{equation}

Here $\alpha $ is the angle of rotation around $OX_{3}$ axis; the origin of
the Cartesian coordinate system $\left\{ x_{1},x_{2},x_{3}\right\} $ is
located in the midpoint of the segment $R$, and the axis $OX_{3}$ is
directed from tne centre 1 to the centre 2.

Consider the explicit form of the differential equations resulting from the
procedure of the separation of variables in Eq. (\ref{f2}) in elongated
spheroidal coordinates (\ref{f3}). By presenting the wave function $\Psi
\left( \xi ,\eta ,\alpha ;R\right) $ as a product $F\left( \xi ;R\right)
\cdot G\left( \eta ;R\right) \cdot \Phi \left( \alpha \right) $ and
substituting it into (\ref{f2}) one obtains three ordinary differential
equations, linked by the separation constants $\lambda $ and $m$:
\begin{equation}
\left[ \frac{\partial }{\partial \xi }\left( \xi ^{2}-1\right) \frac{
\partial }{\partial \xi }+a\xi +\left( p^{2}-\gamma \xi ^{2}\right) \left(
\xi ^{2}-1\right) -\frac{m^{2}}{\left( \xi ^{2}-1\right) }+\lambda \right]
F\left( \xi ;R\right) =0,  \label{f4a}
\end{equation}

\begin{equation}
\left[ \frac{\partial }{\partial \eta }\left( 1-\eta ^{2}\right) \frac{%
\partial }{\partial \eta }+b\eta +\left( p^{2}-\gamma \eta ^{2}\right)
\left( 1-\eta ^{2}\right) -\frac{m^{2}}{\left( 1-\eta ^{2}\right) }-\lambda
\right] G\left( \eta ;R\right) =0,  \label{f4b}
\end{equation}

\begin{equation}
\left[ \frac{\partial ^{2}}{\partial \alpha ^{2}}+m^{2}\right] \Phi \left(
\alpha \right) =0.  \label{f4c}
\end{equation}

Here we use the notations

\begin{center}
$p=\frac{R}{2}\sqrt{2E^{\prime }}$, \qquad $E^{\prime }=E-\frac{\omega
^{2}R^{2}}{2}$, $\qquad \gamma =\frac{\omega ^{2}R^{4}}{4}$,

$a=\left( Z_{1}+Z_{2}\right) \cdot R$, $\qquad b=\left( Z_{2}-Z_{1}\right)
\cdot R$.
\end{center}

In order to have the complete wave function $\Psi \left( \overrightarrow{r}%
;R\right) $ normalized, the functions $F\left( \xi ;R\right) $ and $G\left(
\eta ;R\right) $ should obey the boundary conditions
\begin{equation}
\left| F\left( 1;R\right) \right| <\infty, \qquad F\left( \infty ;R\right)
<\infty  \label{f5a}
\end{equation}

\begin{equation}
\left| \Phi \left( \pm 1;R\right) \right| <\infty .  \label{f5b}
\end{equation}

The procedure of obtaining the energy terms $E\left( R\right) $ is reduced
to the following steps. At first two boundary problems are considered
independently: (\ref{f4a}), (\ref{f5a}) for the quasiradial and (\ref{f4b}),
(\ref{f5b}) for the quasiangular equations, $\lambda ^{\left( \xi \right) }$
and $\lambda ^{\left( \eta \right) }$ being considered the eigenvalues and $p
$ being left a free parameter. Each of the eigenfunctions can be
conveniently characterized by two quantum numbers $n,$ $m$ and the
eigenvalue $\lambda $, namely: $n_{\xi },$ $m,$ $\lambda ^{\left( \xi
\right) }$ for $F_{n_{\xi ,}m}\left( \xi ;R\right) $ and $n_{\kappa },$ $m,$
$\lambda ^{\left( \xi \right)}$ for $G_{n_{\eta },m}\left( \eta ;R\right) $.
The quantum numbers $n_{\xi },$ $n_{\eta }$ are non-negative integers 0,1,2$%
\ldots $ and coincide with the number of nodes for $F_{n_{\xi ,}m}\left( \xi
;R\right) $, $G_{n_{\eta },m}\left( \eta ;R\right) $ functions on the radial
($1\leq \xi <\infty )$ and angular ($-1\leq \eta \leq 1)$ intervals,
respectively. The general theory of Sturm-Liuville-type one-dimensional
boundary problems implies that the quantum numbers $n_{\xi },$ $n_{\eta },$ $%
m$, remain constant at the continuous variation of the intercentre distance $%
R$, and the eigenvalues $\lambda _{n_{\xi }m}^{\left( \xi \right) }\left(
p,a,\gamma \right) $ or $\lambda _{n_{\kappa }m}^{\left( \eta \right)
}\left( p,b,\gamma \right) $ are non-degenerate.

The pair of one-dimensional boundary problems for $F_{n_{\xi
,}m}\left( \xi ;R\right) $and $G_{n_{\eta },m}\left( \eta
;R\right) $ is equivalent to the initial $eZ_{1}Z_{2}\omega $
problem under condition of equality of the eigenvalues $\lambda
_{n_{\xi }m}^{\left( \xi \right) }\left( p,a,\gamma \right)
=\lambda _{n_{\kappa }m}^{\left( \eta \right) }\left( p,b,\gamma
\right)$ and the account of $p,$ $a,$ $b,$ $\gamma$ relationship
with the $E,Z_{1},Z_{2},\omega,R$ \quad parameters. The
eigenvalues \quad $E_{n_{\xi }n_{\eta }m}$, $\lambda
_{n_{\xi}n_{\eta}m}$ \quad and \quad eigenfunctions \quad
$\Psi_{n_{\xi}n_{ \eta}m}\left( \overrightarrow{r};R\right)$ of
the three-dimensional $ eZ_{1}Z_{2}\omega$ problem are enumerated
by a set of quantum numbers $j=\left( n_{\xi }n_{\eta }m\right) $
which are conserved at the continuous variation of $Z_{1},$
$Z_{2},$ $\omega ,$ $R$ parameters:
\begin{equation}
E_{j}\left( R\right) =E_{n_{\xi }n_{\eta }m}\left( R,Z_{1},Z_{2},\omega
\right) ,  \label{f6}
\end{equation}

\begin{equation}
\Psi _{j}\left( \overrightarrow{r};R\right) =N_{j}\left( R\right) \cdot
F\left( \xi ;R\right) \cdot G\left( \eta ;R\right) \cdot \frac{e^{im\alpha}
}{\sqrt{2\pi }}  \label{f7}
\end{equation}

The normalization constant $N_{j}\left( R\right) $ is found from the
condition

\begin{equation}
{\int d\Omega }\Psi _{i}^{*}\Psi _{j}=\delta _{ij}, \qquad d\Omega =\frac{%
R^{3}}{8}\left( \xi ^{2}-\eta ^{2}\right) d\xi d\eta d\alpha =\frac{R^{3}}{8}%
d\tau d\alpha,  \label{f8}
\end{equation}
where $\delta _{ij}$ is the Kronecker symbol, and $\Omega =\left\{ \xi ,\eta
,\alpha \mid 1\leq \xi <\infty ,-1\leq \eta \leq 1,0\leq \alpha <2\pi
\right\} $. Hence, the system of functions $\left\{ \Psi _{j}\left(
\overrightarrow{r};R\right) \right\} $ forms a complete set of
orthonormalized wave functions.

Now we proceed to establish the relationship between the symmetry properties
of the $eZ_{1}Z_{2}\omega $ problem and the above separation of variables in
the Schroedinger equation (\ref{f2}) in the elongated spheroidal coordinates
(\ref{f3}). The very fact of such separation indicates an additional (with
respect to the geometrical one).symmetry of the Hamiltonian (\ref{f2}),
causing the existence of an additional integral of motion, whose operator
commutates with $\widehat{H}$ and the operator $\widehat{L}_{3}$, the
projection of the angular moment on the intercentral axis $\overrightarrow{R}
$. In order to reveal it, we exclude the energy parameter $p^{2}$ and the
magnetic quantum number $m$ from the above differential equation system (\ref
{f4a})-(\ref{f4c}). Thus we derive at the equation
\begin{equation}
\widehat{\lambda }\Psi _{j}\left( \overrightarrow{r};R\right) =\lambda
_{j}\Psi _{j}\left( \overrightarrow{r};R\right) ,  \label{f9}
\end{equation}
where $\widehat{\lambda }$ denotes a differential operator
\begin{eqnarray}
\widehat{\lambda }=\frac{1}{\xi ^{2}-\eta ^{2}}\left\{ \left( \xi
^{2}-1\right) \frac{\partial }{\partial \eta }\left( 1-\eta
^{2}\right) \frac{\partial }{\partial \eta }-\left( 1-\eta
^{2}\right) \frac{\partial }{
\partial \xi }\left( \xi ^{2}-1\right) \frac{\partial }{\partial \xi }
\right\} +  \nonumber \\ +\left[ \frac{1}{1-\eta
^{2}}-\frac{1}{\xi ^{2}-1}\right] \frac{\partial ^{2}}{\partial
\alpha ^{2}}-RZ_{1}\frac{\xi \eta +1}{\xi +\eta }+RZ_{2}\frac{ \xi
\eta -1}{\xi -\eta }+\frac{\omega ^{2}R^{4}}{4}\left( \xi
^{2}-1\right) \left( 1-\eta ^{2}\right) .  \label{f10}
\end{eqnarray}

The separation constant $\lambda _{j}$ is the eigenvalue of this operator,
and the solutions of Eq. (\ref{f2}) are its eigenfunctions. Since in the
limit $\omega \rightarrow 0$ the model $eZ_{1}Z_{2}\omega $ problem is
reduced to the problem of two purely Coulomb centres $eZ_{1}Z_{2}$ \cite{10}%
, it is $a$ $priori$ obvious that the operator $\widehat{\lambda }$ should
be a linear combination of the operators $\widehat{L}_{3}$, $\widehat{P}%
_{3}^{2}$ and $\widehat{H}$ (here $L$ is the orbital moment operator and $%
\widehat{P}_{3}$ is the third component of the momentum) which in the
considered limit is reduced to the operator of the separation constant for
the $eZ_{1}Z_{2}$ problem \cite{10}. To determine the weight factors and the
free constant in the mentioned linear combination we compare the expression (%
\ref{f10}) with the explicit form of the operators $\widehat{L}_{3}$, $%
\widehat{P}_{3}^{2}$ and $\widehat{H}$ in the elongated spheroidal
coordinates (\ref{f3}). After simple but rather dull calculations we finally
obtain the algebraic expression for the separation constant operator in the $%
eZ_{1}Z_{2}\omega $ problem:
\begin{equation}
\widehat{\lambda }=-\widehat{L}^{2}+x_{3}R\left( \frac{Z_{2}}{r_{2}}-\frac{%
Z_{1}}{r_{1}}\right) -\omega ^{2}R^{2}\left( x_{3}^{2}+\frac{R^{2}}{4}%
\right) +\frac{R^{2}}{4}\left( 2\widehat{H}-\widehat{P}_{3}^{2}\right) .
\label{f11}
\end{equation}

The fact the operator $\widehat{\lambda }$ commutating with the Hamiltonian $%
\widehat{H}$ and the operator $\widehat{L}_{3}$ of the angular moment
projection onto the intercentral axis $\overrightarrow{R}$, can be easily
verified by direct calculations of commutational relations $\left[ \widehat{%
H },\widehat{\lambda }\right] =\left[ \widehat{\lambda },\widehat{L}
_{3}\right] =0$. Thus, the operators $\widehat{H}$, $\widehat{L}_{3}$, $%
\widehat{\lambda }$ have a common complete system of eigenfunctions and can
be diagonalized simultaneously. The given representation corresponds to the
separation of variables in Eq. (\ref{f2}) in the elongated spheroidal
coordinates (\ref{f3}): the general eigenfunction of the operators $\widehat{%
H} $, $\widehat{L}_{3}$, $\widehat{\lambda }$ is described as a product (\ref
{f7}).

The purely geometric symmetry group of the Hamiltonian $eZ_{1}Z_{2}\omega $
is the $O_{2}$ group containing rotations around the intercentral axis $%
\overrightarrow{R}$ and reflections in the planes containing this axis. In
the symmetrical case ($Z_{1}=Z_{2}=Z$) the $eZZ\omega $ system possesses an
additional element of geometrical symmetry - the reflection in the plane,
perpendicular to the $\overrightarrow{R}$ vector and cutting it at its
centre.

In addition to the geometrical symmetry, the $eZ_{1}Z_{2}\omega $ problem
possesses higher dynamic symmetry, related to the exact separation of
variables in the Schroedinger equation (\ref{f2}) in the elongated
spheroidal coordinates (\ref{f3}).

In the following subsections we show how, by means of the separation of
variables method, the dynamic symmetry group of the quantum-mechanical
problem $eZ_{1}Z_{2}\omega $ can be determined.

\section{The representations of the group $P\left( 3\right) \otimes P\left(
2,1\right)$}

Consider a group $P\left( 3\right) \otimes P\left( 2,1\right) $ being a
direct product of two groups of motions of three-dimensional spaces $P\left(
3\right) $ and $P\left( 2,1\right) .$

We remind that the group $P\left( 3\right) $ (known also as the Galilean
group $E\left( 3\right) $) consists of displacements (translations) and
rotations (revolutions) of the Euclidian space of coordinates $y_{i}$ with a
metric
\begin{equation}
y_{i}y_{i}=y_{1}^{2}+y_{2}^{2}+y_{3}^{2}, \quad i=1,2,3.  \label{f12}
\end{equation}

Here and below the twice repeated indices imply summation.

The $P\left( 2,1\right) $ group (the Poincare group of the three-dimensional
space, denoted also as $E\left( 2,1\right) $) consists of translations and
rotations of rhe pseudo-Euclidian space of coordinates $y_{\mu }$ with a
metric
\begin{equation}
y_{\mu }y_{\mu }=y_{4}^{2}+y_{5}^{2}-y_{6}^{2}, \quad \mu =4,5,6.
\label{f13}
\end{equation}

The infinitesimal generators of the $P\left( 3\right) $ group
\begin{equation}
x_{j}=-i\frac{\partial }{\partial y_{j}}, \qquad \pounds _{jk}=-i\left( y_{j}%
\frac{\partial }{\partial y_{k}}-y_{k}\frac{\partial }{ \partial y_{j}}%
\right), \qquad j,k=1,2,3  \label{f14}
\end{equation}
and of the $P\left( 2,1\right) $ group
\begin{eqnarray}
x_{\mu } &=&-i\frac{\partial }{\partial y_{\mu }}, \mu =4,5,6; \pounds
_{46}=-i\left( y_{4}\frac{\partial }{\partial y_{6}}+y_{6}\frac{ \partial }{%
\partial y_{4}}\right), \pounds _{56}=-i\left( y_{5}\frac{ \partial }{%
\partial y_{6}}+y_{6}\frac{\partial }{\partial y_{5}}\right) ,  \nonumber \\
\pounds _{45} &=&-i\left( y_{4}\frac{\partial }{\partial y_{5}}-y_{5}\frac{
\partial }{\partial y_{4}}\right)  \label{f15}
\end{eqnarray}
can be easily veified to satisfy the known structure relations:

\[
\left[ x_{i},x_{j}\right] =0,\left[ x_{i},\pounds _{jk}\right] =i\left(
\delta _{ik}x_{j}-\delta _{ij}x_{k}\right) ,\delta _{ij}=\left\{
\begin{array}{cc}
1, & i=j=1,2,3 \\
0, & i\neq j
\end{array}
,\right\}
\]

\[
\left[ \pounds _{12},\pounds _{23}\right] =i\pounds _{31,} \qquad \left[
\pounds _{31},\pounds _{12}\right] =i\pounds _{23,} \qquad \left[ \pounds
_{23},\pounds _{31}\right] =i\pounds _{12,}
\]

\[
\left[ x_{\mu },x_{\nu }\right] =0,\left[ x_{\sigma },\pounds _{\mu \nu
}\right] =i\left( \delta _{\sigma \nu }x_{\mu }-\delta _{\sigma \mu }x_{\nu
}\right) ,\delta _{\mu \nu }=\left\{
\begin{array}{cc}
1, & \mu =\nu =4,5 \\
-1, & \mu =\nu =6 \\
0, & \mu \neq \nu
\end{array}
,\right\}
\]

\[
\left[ \pounds _{46},\pounds _{56}\right] =i\pounds _{45,} \qquad \left[
\pounds _{56},\pounds _{45}\right] =i\pounds _{46,} \qquad \left[ \pounds
_{45},\pounds _{46}\right] =i\pounds _{56,}
\]

\begin{equation}
\left[ x_{i},x_{\mu }\right] =0, \qquad \left[ \pounds _{ij},\pounds _{\mu
\nu }\right] =0, \qquad \left[ x_{i},\pounds _{\mu \nu }\right] =0, \qquad
\left[ x_{\mu },\pounds _{ij}\right] =0.  \label{f16}
\end{equation}

Here the indices $i,j,k$ are 1, 2, 3, and $\mu ,\nu ,\sigma $ are $4,5,6$.
Note that in order to simplify the notations the ''\symbol{94}'' symbol over
the operators is omitted since in this context no threat of ambigiuty can
arise.

The differential operators (\ref{f14}), (\ref{f15}) act in the space of
functions $f_{j}\left( \overrightarrow{y}\right) $ which depend on the
choice of the complete set of the diagonal operators in $P\left( 3\right)
\otimes P\left( 2,1\right) $. Here $j$ is the set of the eigenvalues of
these operators. It is worth notice that in the chosen repreentation the
functions $f_{j}\left( \overrightarrow{y}\right) $ are scalar. In the
general case the generators (\ref{f14}), (\ref{f15}) can possess the spin
part, and $f_{j}\left( \overrightarrow{y}\right) $ can be spinors, vectors,
tensors, respectively.

By Fourier transformation
\begin{equation}
f_{j}\left( \overrightarrow{y}\right) =\int \exp \left( -i\overrightarrow{x}
\overrightarrow{y}\right) \Psi _{j}\left( \overrightarrow{x}\right) d
\overrightarrow{x}  \label{f17}
\end{equation}
we proceed to the x-representation and choose in the $P\left( 3\right)
\otimes P\left( 2,1\right) $ group such set of diagonal intercommutating
operators:
\begin{equation}
\widehat{C}_{1}=x_{i}x_{i}, \qquad \widehat{C}_{2}=x_{i}x_{i}
\overrightarrow{L}^{2}-x_{i}x_{j}\pounds _{ik}\pounds _{jk},  \label{f18a}
\end{equation}

\begin{equation}
\widehat{C}_{3}=x_{\mu }x_{\mu }, \qquad \widehat{C}_{4}=x_{\mu }x_{\mu }%
\overrightarrow{M}^{2}-x_{\mu }x_{\sigma }\pounds _{\nu \mu }\pounds
_{\sigma \mu },  \label{f18b}
\end{equation}

\begin{equation}
\pounds _{12}, \qquad \pounds _{45}  \label{f18c}
\end{equation}

\begin{equation}
\widehat{E}=-\frac{1}{2\left( x_{6}^{2}-x_{3}^{2}\right) }\left[ -
\overrightarrow{M}^{2}+2\overline{a}x_{6}-\overrightarrow{L}^{2}+2\overline{%
b }x_{3}-4\omega ^{2}\left( x_{6}^{4}-x_{3}^{4}\right) \right] ,
\label{f18d}
\end{equation}

\begin{eqnarray}
\widehat{\lambda } &=&\frac{\left( x_{3}^{2}-\widehat{C}_{1}\right) }{\left(
x_{6}^{2}-x_{3}^{2}\right) }\left[ -\overrightarrow{M}^{2}+2\overline{a}%
x_{6}+4\omega ^{2}\left( \widehat{C}_{1}^{2}-x_{6}^{4}\right) \right] +
\nonumber \\
&&+\frac{\left( x_{6}^{2}-\widehat{C}_{1}\right) }{\left(
x_{6}^{2}-x_{3}^{2}\right) }\left[ -\overrightarrow{L}^{2}+2\overline{b}%
x_{3}-4\omega ^{2}\left( \widehat{C}_{1}^{2}-x_{3}^{4}\right) \right] .
\label{f18e}
\end{eqnarray}

Here
\[
\overrightarrow{L}^{2}=\pounds _{12}^{2}+\pounds _{32}^{2}+\pounds
_{31}^{2}, \qquad \overrightarrow{M}^{2}=\pounds _{46}^{2}+\pounds
_{56}^{2}-\pounds _{45}^{2},
\]
$\omega $, $\overline{a}$ and $\overline{b}$ are constants. Note
that summation over the indices $i,j,k$ is performed according to
the metric (\ref
{f12}), and over the indices $\mu ,\nu ,\sigma -$ according to the metric (%
\ref{f13}).

The introduced operators (\ref{f18a})-(\ref{f18e}) possess inportant
properties. The operators $\widehat{C}_{1}$, $\widehat{C}_{2}$ are the
Casimir operators of $P\left( 3\right) $ group, and the operators $\widehat{C%
}_{3}$, $\widehat{C}_{4}-$ are the Casimir operators of $P\left( 2,1\right) $
group. One can verify by direct calculations that the operators $\widehat{C}%
_{2}$ and $\widehat{C}_{4}$ are equal to zero: $\widehat{C}_{2}=\widehat{C}%
_{4}=0$. This, in turn, means that the considered representation is
degenerate. Further $\pounds _{12},$ $\pounds _{45}-$ are the invariants of
uniparametric subgroups of rotations in $P\left( 3\right) $ and $P\left(
2,1\right) $, respectively, and $\widehat{E}$, $\widehat{\lambda }-$
non-canonical diagonal operators.

By substituting the expression for $\overrightarrow{L}^{2}-2\overline{b}
x_{3}-4\omega ^{2}x_{3}^{4}$ (or $\overrightarrow{M}^{2}-2\overline{a}
x_{6}+4\omega ^{2}x_{6}^{4}$) from (\ref{f18d}) into (\ref{f18e}) can be
given by
\begin{equation}
\widehat{\lambda }=-\overrightarrow{L}^{2}+2\left( \widehat{C}
_{1}-x_{3}^{2}\right) \left[ \widehat{E}-2\omega ^{2}\left( \widehat{C}
_{1}+x_{3}^{2}\right) \right] +2\overline{b}x_{3},  \label{f19a}
\end{equation}
or also
\begin{equation}
\widehat{\lambda }=\overrightarrow{M}^{2}+2\left( \widehat{C}
_{1}-x_{6}^{2}\right) \left[ \widehat{E}-2\omega ^{2}\left( \widehat{C}
_{1}+x_{6}^{2}\right) \right] -2\overline{a}x_{6}.  \label{f19b}
\end{equation}

Our further goal is to construct the basis of eigenvectors $\Psi _{j}\left(
\overrightarrow{x}\right) $ in which the complete set of operators (\ref
{f18a})-(\ref{f18e}) is diagonal in $P\left( 3\right) \otimes P\left(
2,1\right) $ group. For this purpose we introduce a new coordinate system in
the x-space:
\begin{equation}
\begin{array}{ccc}
x_{1}=\frac{R}{2}\sqrt{1-\eta ^{2}}\cos \alpha , & x_{2}=\frac{R}{2}\sqrt{%
1-\eta ^{2}}\sin \alpha , & x_{3}=\frac{R}{2}\eta , \\
x_{4}=\frac{R}{2}\sqrt{\xi ^{2}-1}\cos \beta , & x_{5}=\frac{R}{2}\sqrt{\xi
^{2}-1}\sin \beta , & x_{6}=\frac{R}{2}\xi ,
\end{array}
\label{f20}
\end{equation}
where
\begin{equation}
0\leq R<\infty ,\quad 1\leq \xi <\infty ,\quad -1\leq \eta <+1,\quad 0\leq
\alpha ,\beta \leq 2\pi .  \label{f21}
\end{equation}

Having omitted the intermediate calculations, we write the final expressions
for the operators (\ref{f14}), (\ref{f15}) in the new variables (\ref{f20}):
\begin{equation}
\begin{array}{cc}
\pounds _{23}=-i\left( \sqrt{1-\eta ^{2}}\sin \alpha \frac{\partial }{
\partial \eta }-\eta \frac{\cos \alpha }{\sqrt{1-\eta ^{2}}}\frac{\partial }{
\partial \alpha }\right) , &  \\
\pounds _{31}=-i\left( -\sqrt{1-\eta ^{2}}\cos \alpha \frac{\partial }{
\partial \eta }-\eta \frac{\sin \alpha }{\sqrt{1-\eta ^{2}}}\frac{\partial }{
\partial \alpha }\right) , & \pounds _{12}=-i\frac{\partial }{\partial
\alpha }, \\
\pounds _{46}=-i\left( \sqrt{\xi ^{2}-1}\sin \beta \frac{\partial }{\partial
\xi }+\xi \frac{\cos \beta }{\sqrt{\xi ^{2}-1}}\frac{\partial }{\partial
\beta }\right) , &  \\
\pounds _{56}=-i\left( \sqrt{\xi ^{2}-1}\cos \beta \frac{\partial }{\partial
\xi }-\xi \frac{\sin \beta }{\sqrt{\xi ^{2}-1}}\frac{\partial }{\partial
\beta }\right) , & \pounds _{45}=-i\frac{\partial }{\partial \beta }.
\end{array}
\label{f22}
\end{equation}

It is seen from these formulae that the operators $\pounds _{12}$ and $%
\pounds _{45}$, belonging to the complete set of intercommutating operators (%
\ref{f18a})-(\ref{f18e}), in the coordinate system (\ref{f20}) depend only
on the variables $\alpha $ and $\beta $. Hence, we obtain the following
relations:
\begin{equation}
\begin{array}{c}
-i\frac{\partial }{\partial \alpha }\Psi _{j}\left( \xi ,\eta ,R,\alpha
,\beta \right) =m_{j}\Psi _{j}\left( \xi ,\eta ,R,\alpha ,\beta \right) , \\
-i\frac{\partial }{\partial \beta }\Psi _{j}\left( \xi ,\eta ,R,\alpha
,\beta \right) =\widetilde{m}_{j}\Psi _{j}\left( \xi ,\eta ,R,\alpha ,\beta
\right) ,
\end{array}
\label{f23}
\end{equation}
where $m_{j}$, $\widetilde{m}_{j}$ are the eigenvalues of the $\pounds _{12}$
, $\pounds _{45}$, operators, respectively. The common solution of Eqs. (\ref
{f23}) can now be given in the multiplicative form
\begin{equation}
\Psi _{j}\left( \xi ,\eta ,R,\alpha ,\beta \right) =\varphi \left( \xi ,\eta
,R\right) e^{im_{j}\alpha +i\widetilde{m}_{j}\beta }.  \label{f24}
\end{equation}

The rest of the operators from the complete set (\ref{f18a})-(\ref{f18e}) in
the coordinate system (\ref{f20}) with the account of (\ref{f23}) are given
by

\begin{equation}
\widehat{C}_{1}=\frac{R^{2}}{4}, \quad \widehat{C}_{2}=0, \quad \widehat{C}%
_{3}=-\frac{R^{2}}{4}, \quad \widehat{C}_{4}=0,  \label{f25a}
\end{equation}

\begin{eqnarray}
\widehat{E}=-\frac{2}{R^{2}\left( \xi ^{2}-\eta ^{2}\right) }\left[ \frac{%
\partial }{\partial \xi }\left( \xi ^{2}-1\right) \frac{\partial }{\partial
\xi }+a\xi -\frac{\omega ^{2}R^{4}}{4}\xi ^{4}-\frac{\widetilde{m}_{j}^{2}}{%
\xi ^{2}-1}\right] -  \nonumber \\
-\frac{2}{R^{2}\left( \xi ^{2}-\eta ^{2}\right) }\left[ \frac{\partial }{%
\partial \eta }\left( 1-\eta ^{2}\right) \frac{\partial }{\partial \eta }%
+b\eta +\frac{\omega ^{2}R^{4}}{4}\eta ^{4}-\frac{m_{j}^{2}}{1-\eta ^{2}}%
\right],  \label{f25b}
\end{eqnarray}

\begin{eqnarray}
\widehat{\lambda }=-\frac{\left( 1-\eta ^{2}\right) }{\left( \xi
^{2}-\eta ^{2}\right) }\left[ \frac{\partial }{\partial \xi
}\left( \xi
^{2}-1\right) \frac{\partial }{\partial \xi }+a\xi +\frac{\omega ^{2}R^{4}}{4%
}\left( 1-\xi ^{4}\right) -\frac{\widetilde{m}_{j}^{2}}{\xi
^{2}-1}\right] + \nonumber \\ +\frac{\left( \xi ^{2}-1\right)
}{\left( \xi ^{2}-\eta ^{2}\right) }\left[
\frac{\partial }{\partial \eta }\left( 1-\eta ^{2}\right) \frac{\partial }{%
\partial \eta }+b\eta -\frac{\omega ^{2}R^{4}}{4}\left( 1-\eta ^{4}\right) -%
\frac{m_{j}^{2}}{1-\eta ^{2}}\right] ,  \label{f25c}
\end{eqnarray}
note that $a=\overline{a}R$, $b=\overline{b}R$.

Though in order to solve the question concerning the eigenfunctions of the
complete set of operators (\ref{f18a})-(\ref{f18e}) one can use their
explicit form (\ref{f25a})-(\ref{f25c}) , we give the espressions for the
operators (\ref{f19a}), (\ref{f19b}) in the new coordinates as well, since
they will also be used for another purpose:
\begin{equation}
\widehat{\lambda }=\left[ \frac{\partial }{\partial \eta }\left( 1-\eta
^{2}\right) \frac{\partial }{\partial \eta }+b\eta +\left( 1-\eta
^{2}\right) \frac{R^{2}\widehat{E}}{2}-\frac{\omega ^{2}R^{4}}{4}\left(
1-\eta ^{4}\right) -\frac{m_{j}^{2}}{1-\eta ^{2}}\right] ,  \label{f26a}
\end{equation}

\begin{equation}
\widehat{\lambda }=-\left[ \frac{\partial }{\partial \xi }\left( \xi
^{2}-1\right) \frac{\partial }{\partial \xi }+a\xi +\left( \xi ^{2}-1\right)
\frac{R^{2}\widehat{E}}{2}+\frac{\omega ^{2}R^{4}}{4}\left( 1-\xi
^{4}\right) -\frac{\widetilde{m}_{j}^{2}}{\xi ^{2}-1}\right] .  \label{f26b}
\end{equation}

Now we show how, using the separation of variables method, one can
find $\Psi _{j}\left( \xi ,\eta ,R,\alpha ,\beta \right)$
functions which are common eigenfunctions of the operators
(\ref{f23}), (\ref{f25a})-(\ref{f25c}) and (\ref{f26a}),
(\ref{f26b}). The application of this method is based on the
properties of $\widehat{\lambda }$ operator, expressed by Eq.
(\ref{f26a}), (\ref{f26b}). We choose the basis of eigenvectors
$\Psi _{j}$ where the operator $\widehat{\lambda }$ is diagonal:
\begin{equation}
\widehat{\lambda }\Psi _{j}=\lambda _{j}\Psi _{j}  \label{f27a}
\end{equation}
and represent $\Psi _{j}$ in the form of a product
\begin{equation}
\Psi _{j}\equiv \Psi _{j}\left( \xi ,\eta ,R,\alpha ,\beta \right)
=N_{j}\left( R\right) \cdot F_{j}\left( \xi ;R\right) \cdot G_{j}\left( \eta
;R\right) \cdot \frac{e^{im_{j}\alpha +i\widetilde{m}_{j}\beta }}{\sqrt{2\pi
}},  \label{f27b}
\end{equation}
where $\lambda _{j}$ are the eigenvalues of the operator
$\widehat{\lambda}$, and $N_{j}\left( R\right) -$ is a normalizing
factor. After the separation of variables in (\ref{f27a}) a pair
of ordinary differential equations for the unknown functions
$F_{j}\left( \xi ;R\right) ,$ $G_{j}\left( \eta ;R\right) $, is
obtained:
\begin{equation}
\left[ \frac{\partial }{\partial \xi }\left( \xi ^{2}-1\right)
\frac{\partial }{\partial \xi }+a\xi +\frac{R^{2}E_{j}}{2}\left(
\xi ^{2}-1\right) +\frac{\omega ^{2}R^{4}}{4}\left( 1-\xi
^{4}\right) +\lambda _{j}-\frac{\widetilde{m}_{j}^{2}}{\xi
^{2}-1}\right] F_{j}\left( \xi ;R\right) =0, \label{f28a}
\end{equation}

\begin{equation}
\left[ \frac{\partial }{\partial \eta }\left( 1-\eta ^{2}\right) \frac{%
\partial }{\partial \eta }+b\eta +\frac{R^{2}E_{j}}{2}\left( 1-\eta
^{2}\right) -\frac{\omega ^{2}R^{4}}{4}\left( 1-\eta ^{4}\right) -\lambda
_{j}-\frac{m_{j}^{2}}{1-\eta ^{2}}\right] G_{j}\left( \eta ;R\right) =0.
\label{f28b}
\end{equation}

Here $E_{j}$ are the eigenvalues of the operator $\widehat{E}$. Since the
operator $\widehat{\lambda }$ commutates with all the operators (\ref{f18a}%
)-(\ref{f18d}), the eigenfunctions (\ref{f27b}) of the operator $\widehat{%
\lambda } $ are also the eigenfunctions of the operators (\ref{f18a})-(\ref
{f18d}) in the coordinate system (\ref{f20}).

The invariance of the Hamiltonian of the $eZ_{1}Z_{2}\omega $ problem with
respect to the $P\left( 3\right) \otimes P\left( 2,1\right) $ group is now
obvious. Indeed, at $m_{j}=\widetilde{m}_{j}=m$ the system of equations (\ref
{f28a}), (\ref{f28b}) coincides with the system (\ref{f4a})-(\ref{f4c}).
Hence, at given $a,b,\omega ,R,m_{j},\widetilde{m}_{j}$ the determination of
the eigenvalues $E_{j}=E_{j}\left( R\right) $, $\lambda _{j}=\lambda
_{j}\left( R\right) $ and limited in the corresponding ranges (\ref{f21})
eigenfunctions $F_{j}\left( \xi ;R\right) ,$ $G_{j}\left( \eta ;R\right) $
of the complete set of the commutating operators (\ref{f23}), (\ref{f25a})-(%
\ref{f25c}) is reduced at $m_{j}=\widetilde{m}_{j}=m$ to the solution of the
problem, completely equivalent to the quantum-mechanical problem $%
eZ_{1}Z_{2}\omega $. In this case the common eigenfunctions (\ref{f27b}) of
the complete set (\ref{f23}) and (\ref{f25a})-(\ref{f25c}), which comprise
the basis of the degenerate non-canonical representation of $P\left(
3\right) \otimes P\left( 2,1\right) $ group, coincide within the normalizing
factor with the two-centred functions (\ref{f7}) multiplied by $\exp \left(
im\beta \right) $. The expressions (\ref{f25b}), (\ref{f25c}) for the
operators $\widehat{E}$ and $\widehat{\lambda }$ coincide at $m_{j}=%
\widetilde{m}_{j}=m$ with the expressions for the operators of energy $%
\widehat{H}$ and separation constant $\widehat{\lambda }$ (See (\ref{f10}))
in the $eZ_{1}Z_{2}\omega $ problem in the elongated spheroidal coordinate
system (\ref{f3}). The variable $R$, being used to express the Casimir
operators of $P\left( 3\right) \otimes P\left( 2,1\right) $ group, in the $%
eZ_{1}Z_{2}\omega $ problem is equal to the intercentral distance.

The operators (\ref{f23}), (\ref{f25a})-(\ref{f25c}) are Hermitian in the
scalar product
\begin{equation}
\left\langle \Psi _{i}\mid \Psi _{j}\right\rangle ={\int \Psi
_{i}^{*}\Psi _{j}d\overline{\Omega }},  \label{f29}
\end{equation}
where $\overline{\Omega }$ corresponds to the range (\ref{f21}) and the
volume element $d\overline{\Omega }=\xi d\xi \cdot d\eta \cdot d\alpha \cdot
d\beta $. Thus, the representation, corresponding to the set (\ref{f25a})-(%
\ref{f25c}) and (\ref{f26a}), (\ref{f26b}), is unitary.

One of the possible consequences of the above group interpretation of the
solutions of the $eZ_{1}Z_{2}\omega $ problem consists in the calculation of
the matrix elements of generators (\ref{f22}) in the non-canonical basis (%
\ref{f27b}) being reduced to the calculation of the two-centred integrals
over the variable $\xi $ and the similar integrals over the variable $\eta $
. This circumstance is the base for the deduction (without the use of the
explicit form of the solutions of the system of Eqs. (\ref{f4a})-(\ref{f4c})
of a specific linear algebra of two-centred integrals.It consists of a sum
of two independent subalgebras: one - for the radial integrals, containing
polynomials over $\xi $, $\sqrt{\xi ^{2}-1}$ and $\frac{\partial }{\partial
\xi }$, and the other - for the angular integrals, containing polynomials
over $\eta $, $\sqrt{1-\eta ^{2}}$ and $\frac{\partial }{\partial \eta }$.
But in specific quantum-mechanical calculations of the energies and wave
functions of various states of three-quark systems the calculations of
two-centred integrals, containing the derivatives over $R$, are required.
The standard way, resulting in the construction of the algebra of such kind
of integrals, consists in the extension of the semiordinary group $P\left(
3\right) \otimes P\left( 2,1\right) $ to the ordinary one $P\left(
5,1\right) $, being realized by the motions of the six-dimensional
coordinate space $y_{\nu }$ with the metric
\begin{equation}
y_{\nu }y_{\nu }=y_{1}^{2}+y_{2}^{2}+y_{3}^{2}+y_{4}^{2}+y_{5}^{2}-y_{6}^{2}.
\label{f30}
\end{equation}

Having complemented the set of generators (\ref{f14}), (\ref{f15}) by nine
more generators
\begin{equation}
\begin{array}{cc}
\pounds _{j4}=-i\left( y_{j}\frac{\partial }{\partial
y_{4}}-y_{4}\frac{\partial }{\partial y_{j}}\right) , & j=1,2,3,
\\ \pounds _{j5}=-i\left( y_{j}\frac{\partial }{\partial
y_{5}}-y_{5}\frac{\partial }{\partial y_{j}}\right) , & \pounds
_{j6}=-i\left( y_{j}\frac{\partial }{\partial y_{6}}+y_{6}\frac{\partial }{\partial y_{j}}\right),
\end{array}
\label{f31}
\end{equation}
we proceed to the x-representation and choose the set of diagonal
commutating operators, corresponding to the set (\ref{f18a})-(\ref{f18e}).
Additional diagonal operators, arising in the group $P\left( 5,1\right) $
due to the degeneracy of the chosen representation do not result in any new
relations. In the coordinate system (\ref{f20}) we obtain the same equations
(\ref{f28a}), (\ref{f28b}) which are reduced to the problem (\ref{f4a})-(\ref
{f4c}) and whose solutions in the case of the group $P\left( 5,1\right) $
will be realized on the cone
\begin{equation}
x_{1}^{2}+x_{2}^{2}+x_{3}^{2}+x_{4}^{2}+x_{5}^{2}-x_{6}^{2}=0.  \label{f32}
\end{equation}

In this case the generators (\ref{f31}) in the coordinate representation in
the coordinate system (\ref{f20}) are given by
\begin{eqnarray*}
\pounds _{14} &=&-i\sqrt{\left( \xi ^{2}-1\right) \left( 1-\eta ^{2}\right) }%
\cdot [\cos \alpha \cdot \cos \beta \cdot \left( \xi \frac{\partial }{%
\partial \xi }+\eta \frac{\partial }{\partial \eta }\right) -\frac{\cos
\alpha \cdot \sin \beta }{\left( \xi ^{2}-1\right) }\frac{\partial }{%
\partial \beta }+ \\
&&+\frac{\cos \beta \cdot \sin \alpha }{\left( 1-\eta ^{2}\right) }\frac{%
\partial }{\partial \alpha }-R\cos \alpha \cdot \cos \beta \frac{\partial }{%
\partial R}],
\end{eqnarray*}

\begin{eqnarray*}
\pounds _{24} &=&-i\sqrt{\left( \xi ^{2}-1\right) \left( 1-\eta ^{2}\right) }%
\cdot [\sin \alpha \cdot \cos \beta \cdot \left( \xi \frac{\partial }{%
\partial \xi }+\eta \frac{\partial }{\partial \eta }\right) -\frac{\sin
\alpha \cdot \sin \beta }{\left( \xi ^{2}-1\right) }\frac{\partial }{%
\partial \beta }- \\
&&-\frac{\cos \beta \cdot \cos \alpha }{\left( 1-\eta ^{2}\right) }\frac{%
\partial }{\partial \alpha }-R\sin \alpha \cdot \cos \beta \frac{\partial }{%
\partial R}],
\end{eqnarray*}

\begin{eqnarray*}
\pounds _{34} &=&i\sqrt{\xi ^{2}-1}\cdot [\cos \beta \cdot \left( \xi \eta
\frac{\partial }{\partial \xi }-\left( 1-\eta ^{2}\right) \frac{\partial }{%
\partial \eta }\right) -\frac{\eta \sin \beta }{\left( \xi ^{2}-1\right) }%
\frac{\partial }{\partial \beta }-R\eta \cdot \cos \beta \frac{\partial }{%
\partial R}+ \\
&&+\frac{\sin \alpha \cdot \sin \beta }{\left( 1-\eta ^{2}\right) }\frac{%
\partial }{\partial \alpha }-R\cos \alpha \cdot \sin \beta \frac{\partial }{%
\partial R}],
\end{eqnarray*}

\begin{eqnarray*}
\pounds _{25} &=&-i\sqrt{\left( \xi ^{2}-1\right) \left( 1-\eta ^{2}\right) }%
\cdot [\sin \alpha \cdot \sin \beta \cdot \left( \xi \frac{\partial }{%
\partial \xi }+\eta \frac{\partial }{\partial \eta }\right) +\frac{\sin
\alpha \cdot \cos \beta }{\left( \xi ^{2}-1\right) }\frac{\partial }{%
\partial \beta }- \\
&&-\frac{\cos \alpha \cdot \sin \beta }{\left( 1-\eta ^{2}\right) }\frac{%
\partial }{\partial \alpha }-R\sin \alpha \cdot \sin \beta \frac{\partial }{%
\partial R}],
\end{eqnarray*}

\[
\pounds _{35}=-i\sqrt{\xi ^{2}-1}\cdot [\sin \beta \cdot \left( \xi \eta
\frac{\partial }{\partial \xi }-\left( 1-\eta ^{2}\right) \frac{\partial }{%
\partial \eta }\right) +\frac{\eta \cos \beta }{\left( \xi ^{2}-1\right) }%
\frac{\partial }{\partial \beta }-R\eta \cdot \sin \beta \frac{\partial }{%
\partial R}],
\]

\[
\pounds _{16}=-i\sqrt{1-\eta ^{2}}\cdot [\cos \alpha \cdot \left( -\left(
\xi ^{2}-1\right) \frac{\partial }{\partial \xi }-\xi \eta \frac{\partial }{%
\partial \eta }\right) -\frac{\xi \sin \alpha }{\left( 1-\eta ^{2}\right) }%
\frac{\partial }{\partial \alpha }+R\xi \cdot \cos \alpha \frac{\partial }{%
\partial R}],
\]

\[
\pounds _{26}=-i\sqrt{1-\eta ^{2}}\cdot [\sin \alpha \cdot \left( -\left(
\xi ^{2}-1\right) \frac{\partial }{\partial \xi }-\xi \eta \frac{\partial }{%
\partial \eta }\right) +\frac{\xi \cos \alpha }{\left( 1-\eta ^{2}\right) }%
\frac{\partial }{\partial \alpha }+R\xi \cdot \sin \alpha \frac{\partial }{%
\partial R}],
\]

\begin{equation}
\pounds _{36}=-i\cdot [-\eta \left( \xi ^{2}-1\right) \frac{\partial }{%
\partial \xi }+\xi \left( 1-\eta ^{2}\right) \frac{\partial }{\partial \eta }%
+\xi \eta R\frac{\partial }{\partial R}].  \label{f33}
\end{equation}

Finally, consider the basis in the group $P\left( 5,1\right) $ - the group
of motions of the six-dimensional coordinate space $y_{\mu }$ with a metric
\begin{equation}
y_{\mu }y_{\mu }=y_{1}^{2}+y_{2}^{2}+y_{3}^{2}-y_{4}^{2}-y_{5}^{2}+y_{6}^{2}.
\label{f34}
\end{equation}

We introduce the infinitesimal generators of this group
\begin{equation}
\begin{array}{ccc}
x_{j}=-i\frac{\partial }{\partial y_{j}}, & L_{jk}=-i\left( y_{j}\frac{%
\partial }{\partial y_{k}}-y_{k}\frac{\partial }{\partial y_{j}}\right) , &
j,k=1,2,3,6, \\
L_{\mu k}=-i\left( y_{\mu }\frac{\partial }{\partial y_{k}}+y_{k}\frac{%
\partial }{\partial y_{\mu }}\right) , & L_{45}=-i\left( y_{4}\frac{\partial
}{\partial y_{5}}-y_{5}\frac{\partial }{\partial y_{4}}\right) , & \mu =4,5
\end{array}
\label{f35}
\end{equation}
and proceed in the x-representation to a new coordinate system
\begin{equation}
\begin{array}{ccc}
x_{1}=\frac{R}{\sqrt{2}}\sqrt{1-\eta ^{2}}\cos \alpha \cos \gamma , & x_{2}=%
\frac{R}{\sqrt{2}}\sqrt{1-\eta ^{2}}\sin \alpha \cos \gamma , & x_{3}=\frac{R%
}{\sqrt{2}}\eta \cos \gamma , \\
x_{4}=\frac{R}{\sqrt{2}}\sqrt{\xi ^{2}-1}\cos \beta \sin \gamma , & x_{5}=%
\frac{R}{\sqrt{2}}\sqrt{\xi ^{2}-1}\sin \beta \sin \gamma , & x_{6}=\frac{R}{%
\sqrt{2}}\xi \sin \gamma ,
\end{array}
\label{f36}
\end{equation}
where $\alpha ,$ $\beta $ run from 0 to $\pi $, and $\gamma -$ from 0 to $%
\frac{\pi }{2}$. By calculating the expressions for the generators (\ref{f35}%
) in the x-representation in the new coordinates (\ref{f36}), we finally
obtain that $L_{jk}$ $(j,k=1,2,3)$, $L_{56,}$ $L_{46},$ $L_{45}$ have the
same form as $\pounds _{jk}$ $(j,k=1,2,3)$, $\pounds _{56,}$ $\pounds _{46},$
$\pounds _{45}$ in the $P\left( 3\right) \otimes P\left( 2,1\right) $ group
in the coordinate system (\ref{f20}). Now following the above scheme of
constructing the complete set of the commutating operators of Eq. (\ref{f18a}%
)-(\ref{f18e})-type, we obtain at $\cos \gamma =\sin \gamma =\frac{1}{\sqrt{2%
}}$, $\frac{\partial }{\partial \gamma }=0$ a problem, completely equivalent
to the $eZ_{1}Z_{2}\omega $ problem.

A similar consideration of wider groups, e.g. conformal groups of
six-dimensional spaces (\ref{f30}), (\ref{f34}) or a group being a direct
product of two conformal groups of spaces (\ref{f12}) and (\ref{f13}),
results at the choice of the corresponding set of the commutating operators
(of Eq. (\ref{f18a})-(\ref{f18e})-type) to a problem, equivalent to the $%
eZ_{1}Z_{2}\omega $ problem. The calculation of the matrix elements of the
generators of these groups is reduced to the calculation of two-centred
integrals, some of which contain the first and the second derivatives over $R
$.

Without a detailed consideration of all aspects of the chosen
representations of the mentioned groups note only that all these
representations are non-canonical representations in group theory. For
instance, for a group of three-dimensional rotations the non-canonical
representations were first considered in \cite{17} in the context of the
quantum theory of asymmetric rotator.

Note also that the considered groups $P\left( 3\right) \otimes P\left(
2,1\right) $, $P\left( 5,1\right) $, $P\left( 4,2\right) $ are the groups of
motions (translations and rotations) of the corresponding spaces, not the
groups of rotations like in the case of the hydrogen-like atom groups.

\section{Conclusions and final remarks}

Summarizing the results of the work, we focus on its most important points.
By means of the separation of variables method an additional spheroidal
integral of motion $\widehat{\lambda }$ is constructed, whose eigenvalues
are the separation constant in the model quantum-mechanical $%
eZ_{1}Z_{2}\omega $ problem. This has enabled the dynamic symmetry groups of
this problem to be determined and the group properties of its solutions to
be studied. $P\left( 3\right) \otimes P\left( 2,1\right) $, $P\left(
5,1\right) $, and $P\left( 4,2\right) $ groups are considered as such
dynamic groups, among them $P\left( 3\right) \otimes P\left( 2,1\right) $
possessing the smallest number of parameters.

While searching for the eigenfunctions of the complete set of
intercommutating operators in the $P\left( 3\right) \otimes P\left(
2,1\right) $ group in the case of degenerate unitary representations of this
group a problem is shown to occur, quite equivalent to the
quantum-mechanical $eZ_{1}Z_{2}\omega $ problem. In this case the energy
operator $\widehat{E}$ is not the Casimir operator of this group and,
accordingly, does not commutate with all generators of this group. The
operator $\widehat{E}$ and the operator $\widehat{\lambda }$, corresponding
to the ''additional'' operator of the separation constant $\widehat{E}$,
commutating with $\lambda _{j}$, are included into the set of the diagonal
operators, determining the non-canonical basis in the considered group. The
space of the chosen representation of this dynamic group covers the whole
spectrum of the energy values for the two-centre $eZ_{1}Z_{2}\omega $
problem.

The developed group treatment of the model $eZ_{1}Z_{2}\omega $ problem is
related to the group treatment of the tradiational quantum-mechanical
problem of two Coulomb centres $eZ_{1}Z_{2}$ \cite{10}-\cite{16}. But its
consequence is a more rich linear algebra of two-centred integrals, which
contains the corresponding linear algebra of the $eZ_{1}Z_{2}$ problem as a
partial case (i. e. at $\omega =0$). A separate publication will be devoted
to the construction of such algebra while here we only note that the
presence of the both mentioned algebras enables and essentially simplifies
the quantum-mechanical calculations of matrix elements and effective
potentials in the three-body problem with Coulomb and oscillatory
interactions \cite{10}. In particular, the obtained results may appear
useful at the calculations of the energy spectra of QQq-baryons and
QQg-mesons. Note also that the model $eZ_{1}Z_{2}\omega $ problem can at
certain conditions be treated as a step to the solution of a relativized
Schroedinger equation \cite{18} with a two-centred confinement-type
potential (\ref{f1}).

One of the authors (V.Yu.L.) is grateful to the INTAS international
association for partial financial support of this work (Ref. ${\#}$:
INTAS-99-01326).


\newpage

\begin{thebibliography}{99}
\bibitem{1}  Fock V.Z. Phys., - 1935. - \textbf{98} - P.145.

\bibitem{2}  Bargmann V.Z. Phys., -1936. -\textbf{99} - P.576.

\bibitem{3}  Miller W. Symmetry and Separation of Variables.- Mir, Moscow,
1981.

\bibitem{4}  Fushchych V.I., Nikitin A.G. Symmetry of Equations of Quantum
Mechanics.- Nauka, Moscow, 1990.

\bibitem{5}  Dodonov V.V., Manko V.I. Invariants and Correlated States of
Non-Stationary Quantum Systems. Invariants and Evolution of Non-Stationary
Quantum Systems.- Nauka, Moscow, 1987.

\bibitem{6}  Richard J.M.// Phys.Rep. - 1992. - \textbf{212} - P. 1.

\bibitem{7}  Mukhoijee S.N., Nag R., Sanyal S., Morii T., Mori-Shita J.,
Tsuge M.// Phys. Rep. - 1993. - \textbf{231} - P. 203.

\bibitem{8}  Vshivtsev A.S., Prokopov A.V., Sorokin V.N., Tatarintsev A.V.
// Yadernaya Fizika - 1998. - v.\textbf{61}, ${\#}$ 2 - P. 218-222. [In
Russian]

\bibitem{9}  Mur V.D., Popov V.S., Simonov Yu. A., Yurov V.P. // Jh. Eksp.
Teor. Fiz., - 1994. - v.\textbf{105} - P. 3 - 27. [In Russian]

\bibitem{10}  Komarov I.V., Ponomaryov L.I., Slavyanov S.Yu. Spheroidal and
Coulomb Spheroidal Functions.- Nauka, Moscow, 1976.

\bibitem{11}  Coulson C.A., Joseph A. // Int. J. Quant. Chem. - 1967. - P.
337-347.

\bibitem{12}  Alliluyev S.P. // Jh. Eksp. Teor. Fiz - 1966. - v.\textbf{51}
- ${\#}$ 6 - P. 1873 - 1879. [In Russian]

\bibitem{13}  Truskova N.F. Elementary Solutions of the Two - Centre Problem
in Quantum Mechanics and the Representations of Groups. // Preprint of the
Joint Institute for Nuclear Reserch, P2-11554, Dubna, 1978. [In Russian]

\bibitem{14}  Truskova N.F. Crossing of the Potential Curves in the Two -
Centre Problem in Quantum Mechanics and the Representations of , Groups. //
Preprint of the Joint Institute for Nuclear Reserch, P2-11988, Dubna, 1978.
[In Russian]

\bibitem{15}  Truskova N.F. Representations of Noncompact Grups and the Two
- Centre Problem in Quantum Mechanics. // Preprint of the Joint Institute
for Nuclear Reserch, P2-11268, Dubna, 1978. [In Russian]

\bibitem{16}  Truskova N.F. Linear Algebra of Integrals of the Two - Centre
Problem in Quantum Mechanics. // Preprint of the Joint Institute for Nuclear
Reserch, P2-11269, Dubna, 1978. [In Russian]

\bibitem{17}  Lukach I., Smorodinskii Ya.A. Quantum Theory of Non-Symmetry
Rotator and Canonical Representations for Group of the Three-Dimensional
Rotations. // Preprint of the Joint Institute for Nuclear Reserch, P2-7465,
Dubna, 1973. [In Russian]

\bibitem{18}  Atanasov A.A., Marinov A.T. // Yadernaya Fizika - 1998. - v.%
\textbf{61}, ${\#}$ 4 - P. 734-738. [In Russian]
\end{thebibliography}
\end{document}